\documentclass[aps,pra,nobalancelastpage,twocolumn,superscriptaddress]{revtex4-1}

\usepackage{color,amsthm,amsmath,amsxtra,amsfonts,dsfont,graphicx,bm,scalerel, bbm,  graphicx,braket}
\usepackage[colorlinks=true,linkcolor=blue, citecolor=blue, urlcolor=blue, bookmarks]{hyperref}

\newcommand{\be}{\begin{equation}}
\newcommand{\ee}{\end{equation}}
\newcommand{\bea}{\begin{eqnarray}}
\newcommand{\eea}{\end{eqnarray}}

\def\Xint#1{\mathchoice
	{\XXint\displaystyle\textstyle{#1}}%
	{\XXint\textstyle\scriptstyle{#1}}%
	{\XXint\scriptstyle\scriptscriptstyle{#1}}%
	{\XXint\scriptscriptstyle\scriptscriptstyle{#1}}%
	\!\int}
\def\XXint#1#2#3{{\setbox0=\hbox{$#1{#2#3}{\int}$}
		\vcenter{\hbox{$#2#3$}}\kern-.5\wd0}}

\def\dashint{\Xint-}

\theoremstyle{plain}

\begin{document}

\title{Entanglement distribution in the Quantum Symmetric Simple Exclusion Process}

\author{Denis Bernard}
\affiliation{Laboratoire de Physique de l’\'Ecole Normale Sup\'erieure, CNRS,ENS \& PSL University, Sorbonne Universit\'e, Universit\'e de Paris, 75005 Paris, France}

\author{Lorenzo Piroli}
\affiliation{Max-Planck-Institut f\"ur Quantenoptik, Hans-Kopfermann-Str. 1, 85748 Garching, Germany}
\affiliation{Munich Center for Quantum Science and Technology, Schellingstra\ss e 4, 80799 M\"unchen, Germany}

\begin{abstract}
We study the probability distribution of entanglement in the Quantum Symmetric Simple Exclusion Process, a model of fermions hopping with random Brownian amplitudes between neighboring sites. We consider a protocol where the system is initialized in a pure product state of $M$ particles, and focus on the late-time distribution of R\'enyi-$q$ entropies for a subsystem of size $\ell$. By means of a Coulomb gas approach from Random Matrix Theory, we compute analytically the large-deviation function of the entropy in the thermodynamic limit. For $q>1$, we show that, depending on the value of the ratio $\ell/M$, the entropy distribution displays either  two or  three distinct regimes, ranging from low- to high-entanglement. These are connected by points where the probability density features singularities in its third derivative, which can be understood in terms of a transition in the corresponding charge density of the Coulomb gas. Our analytic results are supported by numerical Monte Carlo simulations.
\end{abstract}

\maketitle

\section{Introduction}

Many physical phenomena admit a description in  terms of random variables, whose dynamics is dictated by stochastic processes. While they have been traditionally introduced for open systems, where randomness is acquired through the interaction with the environment~\cite{breuer2002theory}, stochastic processes have recently received renewed attention in  connection with investigations of \emph{typical} features of isolated many-body systems. This trend was driven by the study of random unitary circuits~\cite{nahum2017quantum}, which proved to be ideal toy models to investigate aspects of the dynamics that are notoriously hard to tackle,  including entanglement growth~\cite{nahum2017quantum,rakovszky2019sub,emergent2019zhou,gullans2019entanglement,znidarivc2020entanglement,huang2020dynamics,entanglement2020zhou}, operator spreading~\cite{nahum2018operator,vonKeyserlingk2018operator,chan2018Solution,sunderhauf2018localization,diffusive2018rakovszky,khemani2018operator,hunter2018operator,friedman2019spectral}, dynamical correlations~\cite{friedman2019spectral,kos2021correlations,kos2020chaos}, and scrambling of quantum information~\cite{hosur2016chaos,scrambling2020bertini,piroli2020random}. Similar ideas were also explored in the context of continuous-time Hamiltonian dynamics~\cite{bauer2017stochastic,onorati2017mixing,knap2018entanglement,rowlands2018noisy,zhou2019operator,sunderhauf2019quantum}, and stochastic conformal field theories~\cite{bernard2020entanglement}.

The relevance of stochastic models for generic systems relies on the assumption that the properties of individual random realizations are close to the averaged ones. While this is often a natural expectation, it is typically difficult to obtain quantitative results on the full probability distribution of coherent phenomena such as quantum entanglement~\cite{cotler2020fluctuations,emergent2019zhou,entanglement2020zhou,carollo2019unraveling,carollo2020entanglement}. At the same time, understanding the nature of fluctuations is clearly an important task, and a necessary step towards the generalization of powerful methods developed for classical stochastic systems, such as the well-established Macroscopic Fluctuation Theory~\cite{bertini2005current,bertini2015macroscopic}.

\begin{figure}
\includegraphics[scale=0.98]{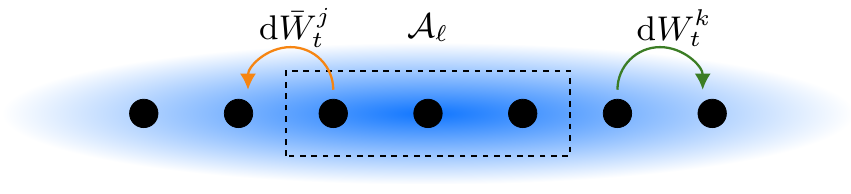}
\caption{Pictorial representation of the Q-SSEP. $M$ fermionic particles, initially in a pure product state, hop with random amplitudes between neighboring sites. We focus on the probability distribution of the entanglement of a subsystem $\mathcal{A}_\ell$, containing $\ell$ neighboring sites. }  
\label{fig:QSSEP}
\end{figure}

Here, we initiate a series of investigations aimed at understanding entanglement fluctuations in a prototypical model for quantum many-body stochastic dynamics: the Quantum Simple Symmetric Exclusion Process (Q-SSEP), cf. Fig.~\ref{fig:QSSEP}. This model, recently introduced in Refs.~\cite{bauer2017stochastic,bauer2019equilibrium}, describes fermions hopping with random amplitudes between neighboring sites, and is particularly useful from the theoretical point of view. On the one hand, given the quadratic form of the Hamiltonian generator, it allows us to employ analytic techniques which are not available in other models. On the other hand, while its mean dynamics reduces to the classical SSEP~\cite{kipnis1989hydrodynamics,eyink1991lattice,derrida2007non,derrida2011microscopic,mallick2015exclusion,bernard2019open,essler2020integrability,duality2020rouven,frassek2020duality}, quantum coherent effects were shown to display a rich phenomenology in this system and its generalizations~\cite{bernard2019open,bernard2020solution,jin2020stochastic}, making it an ideal toy model to build a quantitative understanding of quantum fluctuations. 

We focus on the simplest setting  where the system is initialized in a pure product state, and compute the large-deviation function for the R\'enyi-$q$ entropy of subsystems at late times. Using the Coulomb gas (CG) approach from Random Matrix Theory (RMT), we find that it displays distinct phases, with two of them corresponding to states approaching either a pure state or the maximally mixed one (defining regimes of low and high entanglement, respectively). These regimes are separated by critical points
where the probability density features singularities in its third derivative and which can be understood in terms of a transition in the corresponding charge density of the CG. Our results are supported by numerical Monte Carlo simulations, and open the way towards further studies of fluctuations of entanglement-related quantities in the Q-SSEP, and its generalizations.

The rest of this article is organized as follows. In Sec.~\ref{sec:model} we introduced the Q-SSEP and review previous results on the characterization of the stationary state approached at late times. In Sec.~\ref{sec:coulombd} we lay out the Coulomb-gas approach to the computation of the large-deviation function. We derive a set of equations whose exact solution is presented in Sec.~\ref{sec:exact_solution}. Finally, our conclusions are reported in Sec.~\ref{sec:conclusions}.

\section{The model}\label{sec:model}
We consider a chain of $L$ sites with periodic boundary conditions. The Q-SSEP is formally defined by the Hamiltonian generator 
\be\label{eq:hamiltonian_generator}
d H_{t}=\sum_{j=1}^{L}\left(c_{j+1}^{\dagger} c_{j} d W_{t}^{j}+c_{j}^{\dagger} c_{j+1} d \bar{W}_{t}^{j}\right)\,,
\ee
where $c_j$, $c_j^\dagger$ are canonical fermionic operators, with $\{c_{j}, c_{k}^{\dagger}\}=\delta_{j ,k}$, and $W^j_t$, $\bar{W}^j_t$ are pairs of complex conjugated Brownian motions. They satisfy $\mathrm{d} W_{t}^{j} \mathrm{d} \bar{W}_{t}^{k}=\delta_{j,k} \mathrm{d}t$, and $\mathrm{d} \bar{W}_{t}^{j} \mathrm{~d} W_{t^{\prime}}^{k}=\mathrm{d} W_{t}^{k} \mathrm{~d} \bar{W}_{t^{\prime}}^{j}=0$ for $t\neq t^\prime$, where we used the standard notation in It\^{o} calculus ~\cite{oksendal2003stochastic}. The system is initialized in a pure product state of $M$ particles. Late-time properties turn out to be independent of the specific initial state chosen, but for concreteness we may take $\ket{\Psi(0)}=c^{\dagger}_1\cdots c^\dagger_{M}\ket{0}$, where $\ket{0}$ is the vacuum. We consider the entanglement of a subsystem $\mathcal{A}_{\ell}=\{1,\ldots ,\ell\}$, as measured by the R\'enyi-$q$ entropies 
\be
S_{q}(t)=(1-q)^{-1}\ln{\rm tr}\left[\rho_\ell^q(t)\right]\,,
\ee 
where $\rho_\ell(t)={\rm tr}_{L\setminus\ell}|\Psi(t)\rangle\langle \Psi(t)|$. Clearly, $s_{q}(t)=S_{q}(t)/\ell$ is a stochastic variable distributed according to some probability density $p_{q,t}(s)$, with $0\leq s \leq \ln2$. Our goal is to compute the full distribution of $s_{q}(t)$ for large times, namely $p_{q}(s)=\lim_{t\to\infty} p_{q,t}(s)$, in the limit of large $L$,  $\ell$, $M$, where we fix the ratios  $\xi=\ell/L$, $ m=M/L$.

As a preliminary observation, note that the initial state $\ket{\Psi(0)}$ satisfies Wick's theorem, and its density matrix is completely specified by its covariance matrix $\left(G_0\right)_{i,j}:=\braket{\Psi(0)|c^\dagger_ic_j|\Psi(0)}$. Since the Hamiltonian is quadratic, this remains  true for the evolved state $\ket{\Psi(t)}$, and the system is effectively described by the evolved covariance  matrix $G_t$. The latter also fully determines the value of the R\'enyi entropies $S_q(t)$~\cite{vidal2003entanglement}: denoting by $A^{(\ell)}$ the matrix obtained by selecting the first $\ell$ rows and columns of a matrix $A$, we have 
\be
S_{q}(t)=(1-q)^{-1}\sum_{j=1}^\ell\ln\left[\lambda_j^q+(1-\lambda_j)^q\right]\,,
\ee
where $\{\lambda_j\}_{j=1}^\ell$ are the eigenvalues of $G_t^{(\ell)}$,  satisfying $0\leq \lambda_j\leq 1$. 

In order to make progress, we use that the density $p_q(s)$ satisfies a large-deviation principle; in particular, we will prove that, for $\ell<L$, $\ln p_q(s)\sim -\ell^2 I_q(s)$, for some rate function $I_q(s)$. In this situation, the G\"artner-Ellis theorem applies~\cite{touchette2009large},  stating that $I_q(s)$ can be computed from the knowledge of the cumulant generating function by a Legendre transform:
\be\label{eq:rate_function}
I_q(s)=-\inf _{w}\{w s-f_q(w)\}\,,
\ee
where we introduced $f_q(w)=-\ell^{-2}\ln\mathcal{F}_q(w)$, with $\mathcal{F}_q(w)=\lim_{t\to\infty}\mathbb{E}_t\left[e^{-w\ell S_{q}(t)}\right]$. Writing $S_{q}(t)=\mathcal{S}_{q,\ell}[G_t]$, where we defined the function 
\be
\mathcal{S}_{q,\ell}[G_t]= (1-q)^{-1}{\rm tr}\ln\left[(G_t^{(\ell)})^q+(\openone -G_t^{(\ell)})^q\right]\,,
\ee
we can make use of a result derived in Ref.~\cite{bauer2019equilibrium}, relating large-time expectation values to averages over the unitary group $U(L)$ equipped with the Haar invariant measure. Explicitly, we obtain
\be\label{eq:f_q_haar}
\mathcal{F}_q(w)=\int_{U(L)} {\rm d} \eta(V) \exp\left(-w\ell \mathcal{S}_{q,\ell}\left[G_V\right]\right)\,,
\ee 
with $G_V=V^\dagger G_0V$, and where ${\rm d} \eta(V) $ denotes the Haar measure over $U(L)$. It follows from Eqs.~\eqref{eq:rate_function} and \eqref{eq:f_q_haar} that the problem is reduced to computing the distribution of the subsystem entanglement for a random pure fermionic \emph{Gaussian} state. It is important to stress that  this is different from the analogous problem for Haar random states sampled over the whole many-body space (having dimension $2^{L}$). In that case several exact results were obtained for the full probability distribution of entanglement~\cite{giraud2007distribution,facchi2008phase,nadal2010phase,depasquale2010phase,nadal2011statistical,facchi2013entropy,facchi2019phase}. While we will employ similar techniques, qualitative and quantitative differences arise in our case.

\section{The Coulomb gas approach}\label{sec:coulombd}

The Haar measure over $U(L)$, induces a probability distribution $P[\{\lambda_j\}]$ on the set of eigenvalues of $G_V^{(\ell)}$, which allows us to express Eq.~\eqref{eq:f_q_haar} in the form
\be\label{eq:f_function}
\mathcal{F}_q(w)= \int (\prod_{j=1}^\ell {\rm d}\lambda_j)\, P[\{\lambda_j\}]e^{-w\ell S_q[\{\lambda_j\}]}\,,
\ee 
where $S_q[\{\lambda_j\}]=(1-q)^{-1}\sum_j\ln [\lambda_j^q+(1-\lambda_j)^q]$. For the initial state chosen, simple manipulations give $G^{(\ell)}_V=V^\dagger_{M,\ell} V_{M,\ell}$, where $V_{M,\ell}$ is the $M\times \ell$ sub-matrix containing the first $M$ rows and $\ell$ columns of $V$. Thus, in order to evaluate~\eqref{eq:f_function}, we need the probability distribution induced on the eigenvalues of $V^\dagger_{M,\ell} V_{M,\ell}$, when $V$ is sampled from the Haar invariant measure. It turns out that the latter is known in RMT~\cite{forrester2010log}, and takes the form
\be\label{eq:jacobi_spectrum}
P[\left\{\lambda_{i}\right\}]=\frac{1}{\mathcal{N}}\prod_{j<k }\left|\lambda_{j}-\lambda_{k}\right|^{2} \prod_{i=1}^{\ell} \lambda_{i}^{M-\ell}(1-\lambda_i)^{L-\ell-M}\,,
\ee
where $\mathcal{N}$ is a normalization constant. This distribution defines the so-called $\beta$-Jacobi ensemble (with $\beta=2$), and has been recently exploited for the computation of averaged subsystem entanglement in the context of random non-interacting fermionic ensembles~\cite{liu2018quantum,zhang2020subsytem,lydzba2020eigenstate,lydzba2021entanglement}, see also~\cite{bianchi2021page}. Note that the distribution depends on both $\ell$ and $M$. In the following, we may restrict to $\ell\leq L/2$, since the entanglement for pure states is symmetric under $\ell\mapsto L-\ell$. Furthermore, we may also choose $\ell\leq M$~\footnote{Indeed, if $\ell>M$, we can use that $V_{M,\ell}^\dagger V_{M,\ell}$ has $\ell-M$ zero eigenvalues, and that its non-vanishing eigenvalues coincide with those of $V_{M,\ell}V^\dagger_{M,\ell}$. In this way, we exchanged the role of $\ell$ and $M$: redefining formally $\ell^\prime=M$, $M^\prime=\ell$, $V^\prime=V^\dagger$, we reduced to the case $\ell^\prime<M^\prime$ treated in the main text.}.

\begin{figure}
	\begin{tabular}{c}
		\hspace{-0.2cm}\includegraphics[scale=0.23]{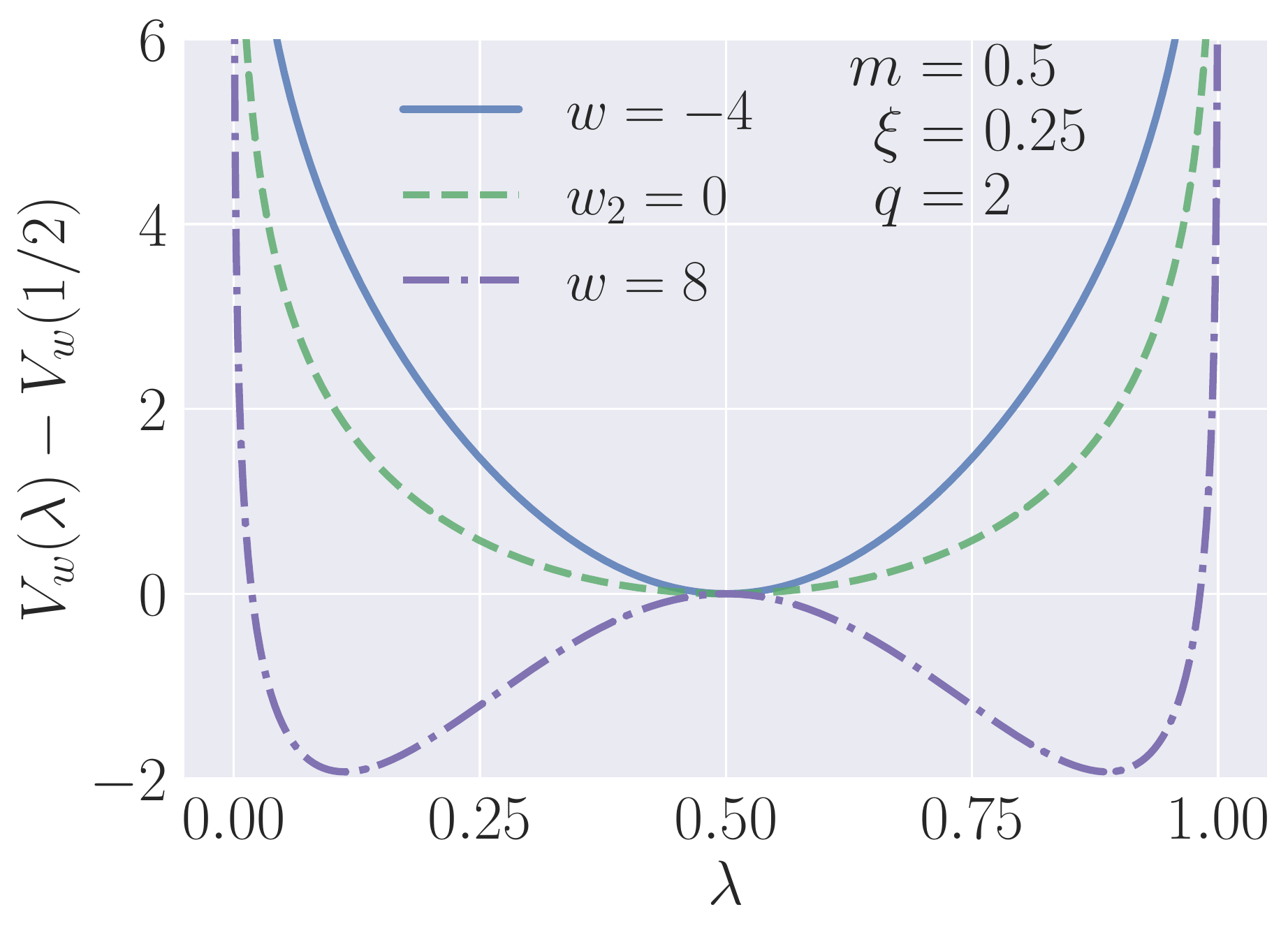}\hspace{-0.2cm}
		\includegraphics[scale=0.23]{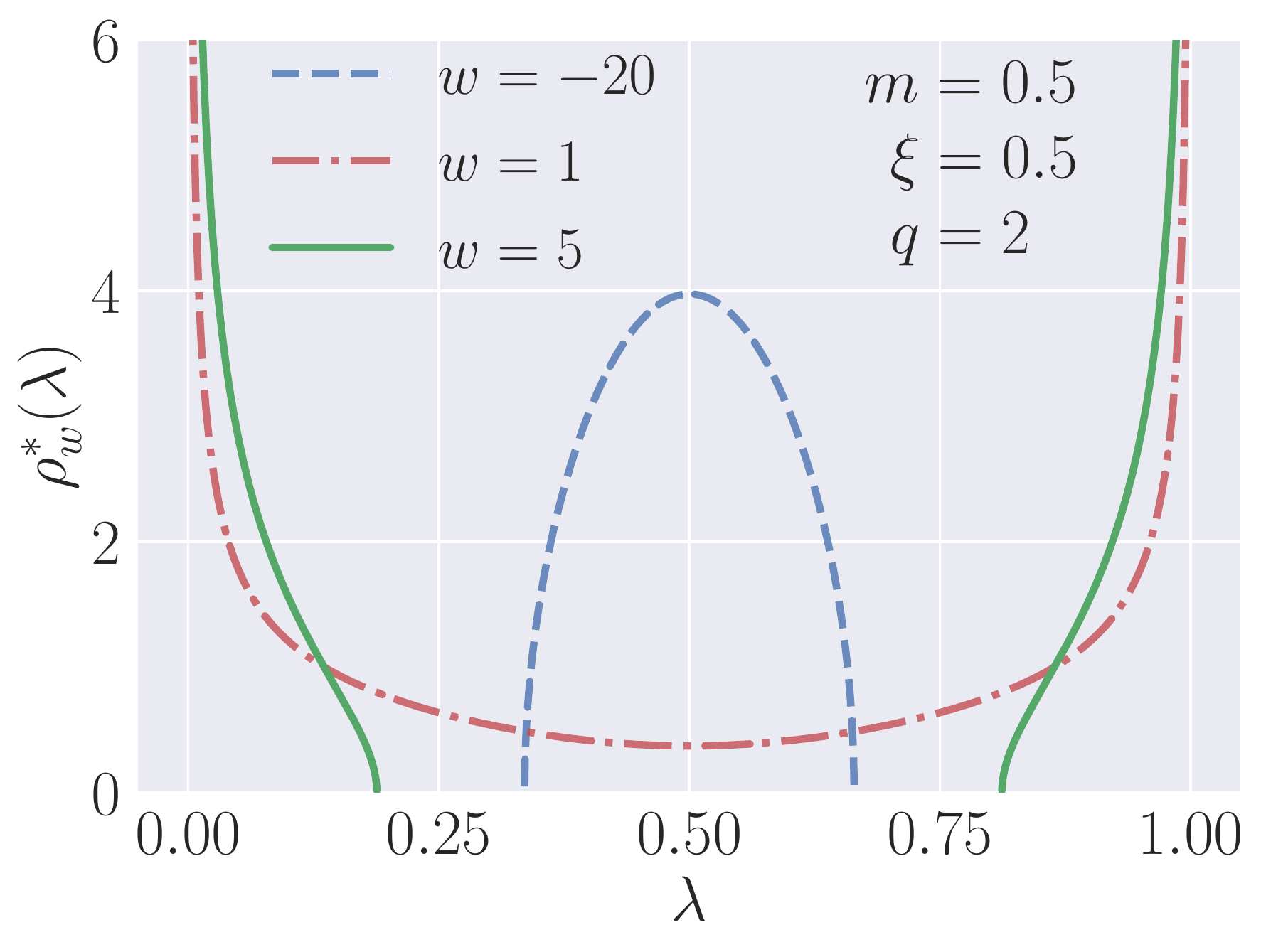}
	\end{tabular}
	\caption{Left: shifted  effective potential $V_{w}(\lambda)$, defined in Eq.~\eqref{eq:effective_potential}, for different values of $w$, and $m=0.5$, $\xi=0.25$, $q=2$. Right: optimal charge distribution for $q=2$, $m=\xi=0.5$. The plot shows $\rho^\ast_w(\lambda)$ for three values of $w$, each corresponding to a different regime.}  
	\label{fig:rho_functions}
\end{figure}

Following Refs.~\cite{nadal2010phase,depasquale2010phase,nadal2011statistical,facchi2013entropy,facchi2019phase}, we use Eq.~\eqref{eq:jacobi_spectrum} as the starting point of our computations, which are based on the CG approach. This is a method routinely applied in RMT, consisting in a mapping between random matrix eigenvalues and  repulsive point charges~\cite{forrester2010log}. The CG analysis of the Jacobi ensemble has been already employed in physical problems, e.g. to study the conductance and the shot noise power for a mesoscopic cavity with two leads~\cite{vivo2008distributions,vivo2010probability}, or to compute the so-called Andreev conductance of a metal-superconductor interface~\cite{damle2011phase}. In order to see how it works,  we rewrite
\be\label{eq:functional}
\mathcal{F}_q(w)=\frac{1}{\mathcal{N}}\int_0^1 (\prod_{j=1}^\ell {\rm d}\lambda_j)\, e^{- \ell^2E_w[\{\lambda_j\}]}\,,
\ee 
with 
\begin{align*}
E_w[\{\lambda_{i}\}]&=-\frac{2}{\ell^2}\sum_{i<j} \ln |\lambda_{i}-\lambda_{j}|
-\frac{(M-\ell)}{\ell^2} \sum_{i} \ln \lambda_{i}
\nonumber\\
-&\frac{(L -M-\ell)}{\ell^2} \sum_{i} \ln \left(1-\lambda_i\right)+wS_q[\{\lambda_{i}\}]/\ell\,.
\end{align*}
Within the CG approach, the function $E_w[\{\lambda_{i}\}]$ is interpreted as the energy of a gas of charged particles with coordinates $\lambda_j\in[0,1]$, which are subject to an external potential. The integral~\eqref{eq:functional} is then understood as a thermal partition function for the CG. In the large-$\ell$ limit, the configuration of the Coulomb charges may be described in terms of the normalized density $\rho(\lambda,\ell)=\ell^{-1}\sum_j\delta(\lambda-\lambda_i)$, and the multiple integral in Eq.~\eqref{eq:functional} can be replaced by an integral over all possible density functions $\rho(\lambda)$, i.e. 
\be\label{eq:functional_integral}
\mathcal{F}_q(w)= \frac{\int \mathcal{D}\rho\, e^{-\ell^2 E_w[\rho]}}{\int \mathcal{D}\rho\, e^{-\ell^2 E_0[\rho]}}\,,
\ee
where the denominator corresponds to the normalization constant $\mathcal{N}$. 
To the leading order in $\ell$~\footnote{In replacing the multiple integral with the functional integral, one needs to take into account the Jacobian $J[\rho]$ associated with the change of coordinates. It can be argued that $J[\rho]\sim e^{O(\ell)}$,  and it may thus be neglected at the leading order in $\ell$~\cite{extreme2008dean}.}, $E_w[\rho]$  reads
\begin{align}
E_w[\rho]=&-\int_0^1 {\rm d}\lambda\int_0^1{\rm d}\mu\, \rho(\lambda)\rho(\mu)\ln |\lambda-\mu|\nonumber\\
+&\int_0^1 {\rm d}\lambda\rho(\lambda)V_w(\lambda)+u\left\{\int_0^1{\rm d}\lambda\,\rho(\lambda)-1\right\}\,,
\end{align}
 where we introduced the Lagrange multiplier $u$ enforcing normalization, and the effective potential
\begin{align}\label{eq:effective_potential}
V_{w}(\lambda)=&-\left(\frac{m}{\xi}-1 \right)\ln \lambda-\left(\frac{1-m}{\xi}-1 \right)\ln\left (1-\lambda\right)\nonumber\\
+&\frac{w}{1-q}\ln\left[\lambda^q+(1-\lambda)^q\right]\,,
\end{align}
where $m$, $\xi$ are the density of fermions and the rescaled interval length introduced before. The functional integrals in~\eqref{eq:functional_integral} may be evaluated by the saddle-point method. This yields $\int \mathcal{D}\rho e^{-\ell^2 E_w[\rho]}\sim e^{-\ell^2E_w[\rho_w^\ast]}$, where $\rho_w^\ast(\lambda)$ is the ``optimal'' charge density, minimizing $E_w[\rho_w^\ast(\lambda)]$ and satisfying 
\be
\int{\rm d}\lambda \rho^\ast_w(\lambda)\ln|\mu-\lambda|=(1/2)V(\mu)+u/2\,.
\ee
Differentiating the last equation with respect to $\mu$ we arrive at
\be\label{eq:integral_equation}
\dashint {\rm d}\lambda\, \frac{\rho^\ast_w(\lambda)}{\mu-\lambda}=\frac{1}{2}V_w^\prime (\mu)\,,
\ee
where $\dashint$ denotes the principal-value integral. Eq.~\eqref{eq:integral_equation} can be formally solved using the so-called Tricomi's  formula~\cite{tricomi1985integral,majumdar2014top}, or a resolvent method~\cite{pipkin1991course,gakhov2014boundary}, which both yield integral representations of the solution which can in general be evaluated numerically. Plugging
\be\label{eq:f_saddle_point}
f_q(w)=E_{w}[\rho^\ast_w]-E_{0}[\rho^\ast_0]\,,
\ee
which is derived from the saddle-point method, into Eq.~\eqref{eq:rate_function}, this finally allows us to obtain a numerical value for $I_q(s)$. In fact, we find that Eq.~\eqref{eq:integral_equation} can be solved analytically for all integers $q>1$. Before discussing the mathematical details, however, it is interesting to observe that its qualitative features can be understood based on the analysis of the CG picture, as we now briefly discuss.

First of all, we note that for $0\leq \xi\leq m\leq 1/2$, the effective potential~\eqref{eq:effective_potential} is always bounded from below. Furthermore, for $w$ negative and with large absolute value, $V_w(\lambda)$ has a single local minimum close to $\lambda=1/2$, cf. Fig.~\ref{fig:rho_functions}. Recalling that $\rho^\ast_w(\lambda)$ describes the distribution of charges subject to the external potential $V_w(\lambda)$, we expect $\rho^\ast_w(\lambda)$ to develop an increasingly sharp peak around this point. This is consistent with our intuition based on the quantum problem: for $w\to-\infty$, maximal entanglement entropies are favored in the average corresponding to $\mathcal{F}_q(w)$, and the most significant states in the average approach the maximally mixed one, i.e.  all the eigenvalues of the covariance matrix should be close to $\lambda=1/2$. For $w$ very large, instead, $V_w(\lambda)$ develops a local maximum close to $\lambda=1/2$, and the Coulomb charges are pushed at the boundaries of $[0,1]$, eventually depleting its central region. Accordingly, we expect $\rho^\ast_w(\lambda)$ to become peaked around $\lambda=0$ and $\lambda=1$, and vanishing in a neighborhood of $\lambda=1/2$. In terms of the quantum problem, this means that all eigenvalues of the covariance matrix are close to $0$ or $1$, i.e. the entanglement vanishes, and we approach a pure state. We will see that the two limits $w\to\pm \infty$ correspond to different phases of the rate function.

\section{The exact solution}\label{sec:exact_solution}

We now present our analytic solution to Eq.~\eqref{eq:integral_equation}. While we were able to obtain explicit expressions for all integers $q>1$, and arbitrary $\xi$ and $m$, they are very cumbersome for general $q$ and $\xi,m<1/2$. For this reason, here we report only the case $q=2$ and $\xi=m=1/2$. Furthermore, we will only present the final result of our analysis, while all the details of our derivations will be reported elsewhere~\footnote{Denis Bernard and Lorenzo Piroli, \emph{in preparation}.}.

In general, we find that $\rho_w^\ast(\lambda)$ displays either two or three distinct phases as a function of $w$, separated by points where $I_q(s)$ develops a discontinuity in its third derivative. Similar kinds of ``third order phase transitions'' are ubiquitous in RMT, appearing  in a wide variety of contexts~\cite{majumdar2014top}. In our case, for $m=\xi=1/2$, and $q=2$, there are three phases, separated by the points $w^\ast_1=-2-\sqrt{2}$ and $w^\ast_2=1+\sqrt{2}$. The first one is characterized by states with large entanglement, and corresponds to $-\infty<w<w^\ast_1$. In this case, $\rho^\ast_w(\lambda)$ has non-zero support over the interval $J^{\rm I}=[\nu_-,\nu_+]\subset[0,1]$, with $\nu_\pm=(1\pm\nu)/2$ and $\nu=-\sqrt{-2w-1}/(w+1)$. It reads \be
\rho_w^\ast(\lambda)=-(2/\pi) (w+1)\frac{ \sqrt{\lambda-\nu _-} \sqrt{\nu _+-\lambda}}{ \lambda^2+(1-\lambda)^2}\,.
\ee
As expected, $\rho^\ast_w(\lambda)$ becomes a delta-function peaked around $\lambda=1/2$ for $w\to-\infty$. Next, for $w^\ast_1<w< w^\ast_2$, we enter a transition regime: $\rho^\ast_w(\lambda)$ has support over the whole interval $J^{\rm II}=(0,1)$, and develops two integral singularities at its boundaries. It reads 
\be
\rho^*_{w}(\lambda)=\frac{1}{\sqrt{\pi^2 \lambda\left(1-\lambda\right)}}g(\lambda)\,,
\ee
with 
\be
g(\lambda)=1+w\{1 - 2^{-1/2}[\lambda^2 + (\lambda - 1)^2]^{-1}\}\,.
\ee
Note that for $w=0$ we recover the spectral density of the Jacobi ensemble, see e.g.~\cite{forrester2010log,forrester2012large,ramli2012spectral}. As $w$ varies from $w_1^\ast$ to $w_2^\ast$, the charge density decreases at the center of the interval, eventually vanishing in $\lambda=1/2$ at $w=w_2^\ast$. Here, we enter the third phase, spanning $w^\ast_2<w<\infty$, which is that of low-entangled states. In this  regime, $\rho^\ast_w(\lambda)$ has non-vanishing support over $J^{\rm III}=(0,\nu_-)\cup (\nu_+,1)$, with $\nu_\pm=(1\pm \nu)/2$ and \be
\nu=\frac{\sqrt{(w-1)^2-2}}{w+1}\,.
\ee 
It has the form 
\be
\rho_{w}^\ast(\lambda)=\frac{|\lambda -1/2|(2+wh(\lambda))}{\pi  \sqrt{(1-\lambda ) \lambda } \sqrt{(1-2 \lambda )^2-\nu (w )^2}}\,,
\ee 
with $h(\lambda)=2-\sqrt{2+2\nu (w )^2}/[1-2 (1-\lambda )\lambda ]$. Importantly, we see that as $w\to\infty$ the support of $\rho^\ast_w(\lambda)$ localizes around $0$ and $1$, yielding vanishing entanglement. We plot the optimal density $\rho^\ast_w(\lambda)$ in Fig.~\ref{fig:rho_functions}, for three values of $w$ corresponding to the phases discussed above.

Let us also mention how this picture is modified when $\xi<1/2$ (and $m=1/2$). In this case, the potential $V_w(\lambda)$ is divergent at $\lambda=0, 1$, and the support of the optimal charge $\rho^\ast_w(\lambda)$ is strictly contained in $[0,1]$. Accordingly, we find that phases I and II merge, so that $\rho^\ast_w(\lambda)$ only displays two phases, separated by the point 
\be
w^\ast(\xi)=\frac{1+2 \sqrt{2} \sqrt{(1-\xi ) \xi }}{2 \xi}\,.
\ee 
The qualitative features of the optimal distributions remain the same, although they do not display singularities at the boundaries of their support for $\xi\neq 1/2$.

\begin{figure}
	\begin{tabular}{c}
		\hspace{-0.2cm}\includegraphics[scale=0.23]{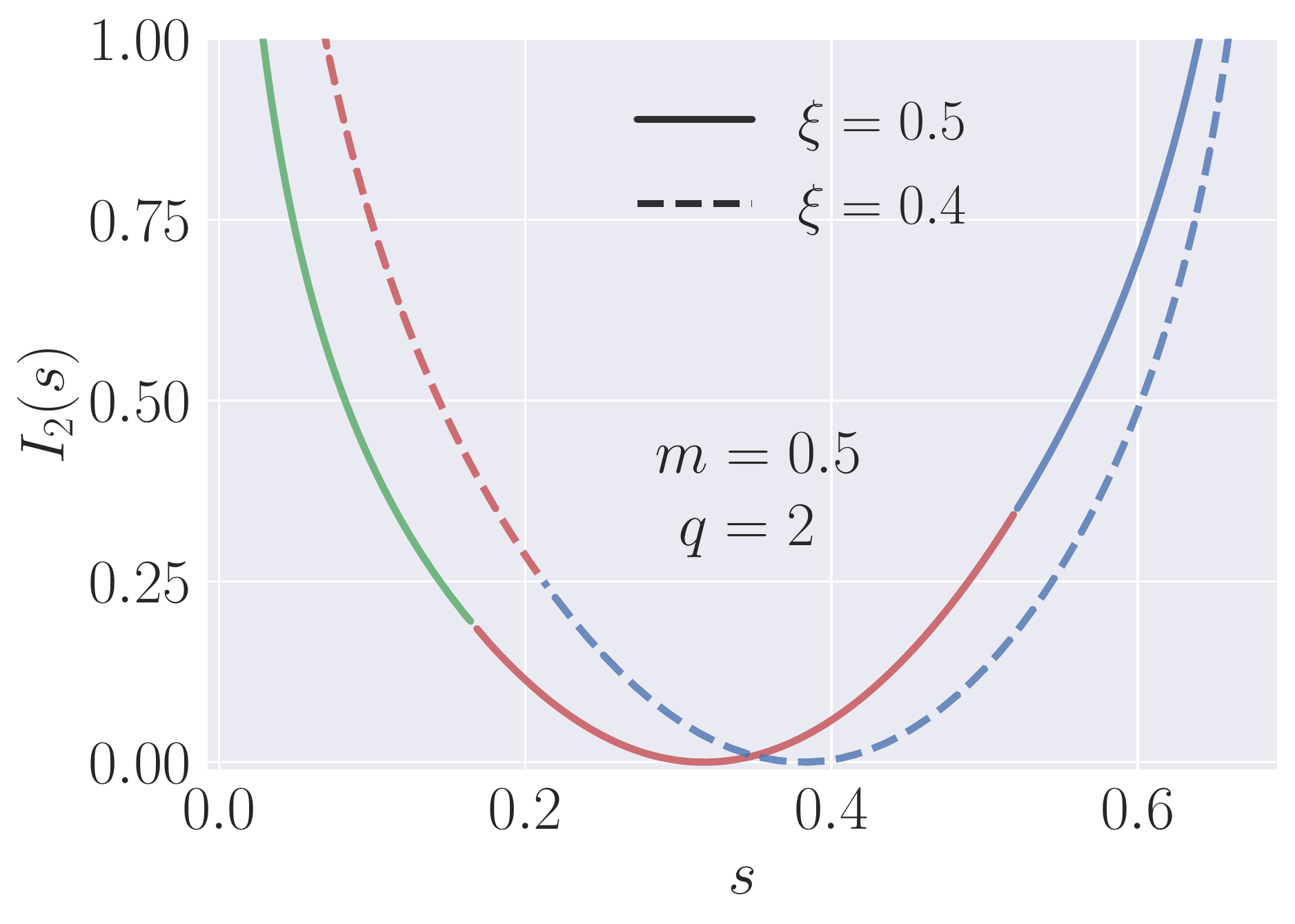}	\hspace{-0.2cm}
		\includegraphics[scale=0.23]{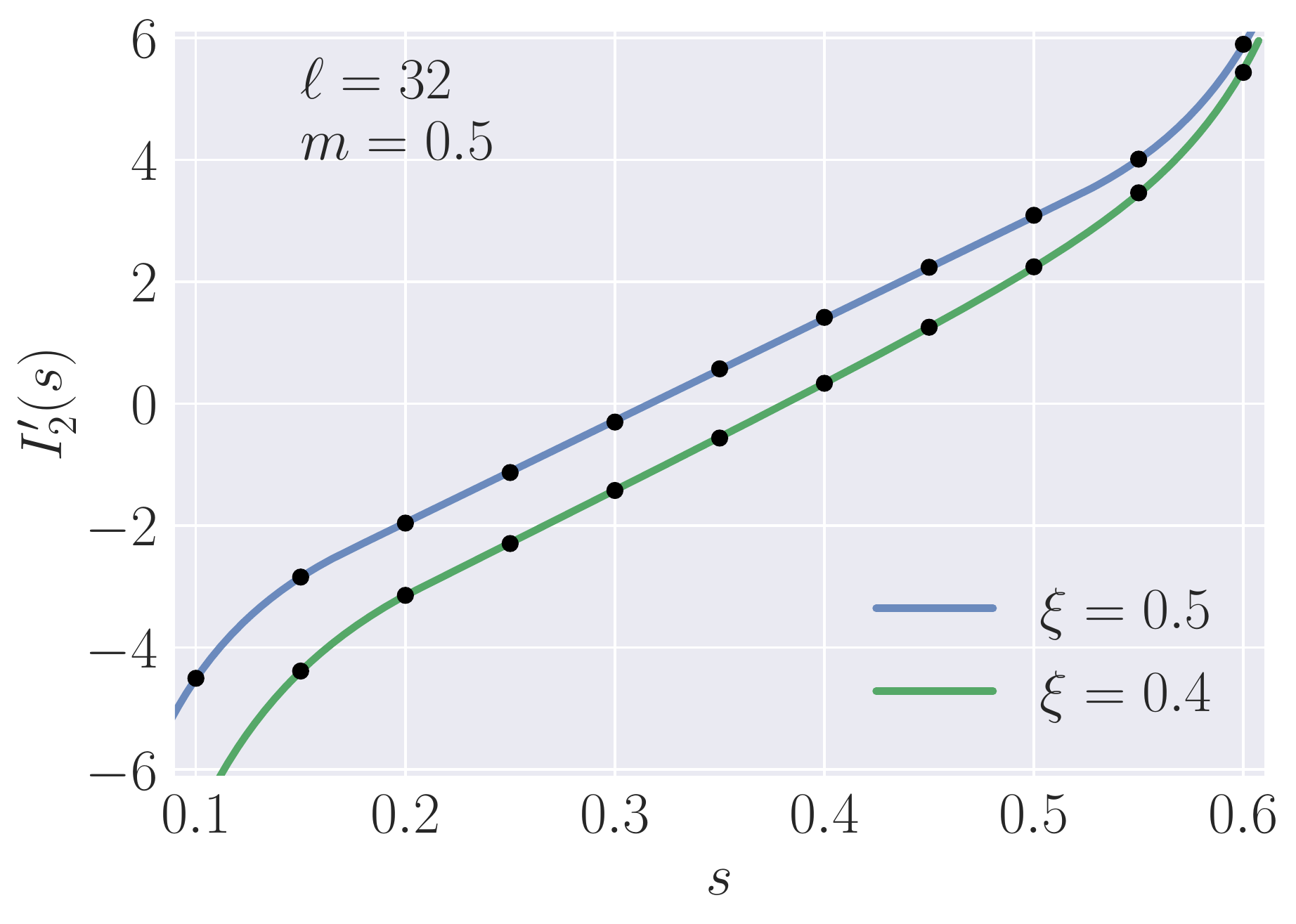}
	\end{tabular}
	\caption{Left: rate function for the R\'enyi-$2$ entropy, for $m=1/2$, and different values of $\xi$. For $\xi<m$ and $\xi=m$ respectively, two and three phases appear, which correspond to different colors. Right: analytic predictions for the derivative $I^\prime_2(s)$ (solid lines), against numerical data from Monte Carlo simulations for $\ell=32$, $L=\ell/\xi$ and $M=L/2$ (dots). The numerical error is not visible at the scales of the plot.}
	\label{fig:rate_renyi2}
\end{figure}

From the knowledge of $\rho_w^\ast(\lambda)$, we can compute the rate function $I_q(s)$. First, it is convenient to rewrite the Legendre transform~\eqref{eq:rate_function} as
\be\label{eq:I_q}
I_q(s)=-w_s s+f_q(w_s)
\ee
where $w_s$ satisfies $df_q(w_s)/dw=s$. Using Eq.~\eqref{eq:f_saddle_point}, and the fact that $\rho_w^\ast(\lambda)$ is the saddle point of $E_w[\rho]$,  this condition is equivalent to $S_q[\rho^\ast_{w_s}] =s$, where 
\be
S_q[\rho]=\int{\rm d}\lambda\, \rho(\lambda)(1-q)^{-1}\ln\left[\lambda^q+(1-\lambda)^q\right]\,.
\ee 
From Eq.~\eqref{eq:I_q} we see that $I_q(s)$ can be computed by evaluating numerically simple integrals~\footnote{In principle, numerical computation of $I_q(s)$  involves a double integral. However, one can get rid of the latter by using the saddle-point equation $\int{\rm d}\lambda \rho^\ast_w(\lambda)\ln|\mu-\lambda|=(1/2)V(\mu)+u/2$, as done in Ref.~\cite{nadal2011statistical}. See~\cite{Note3} for further details.}. We followed this procedure to generate plots of the function $I_q(s)$ for different  values of $\xi$, as reported in Fig.~\ref{fig:rate_renyi2}. As a general feature, we see that the rate function develops singularities at $s=0$, $s=\ln 2$. We also note that we may read off the average value for the entropy, corresponding to the minimum of $I_q(s)$.

To obtain an analytic form for $I_q(s)$, one should invert the relation $S_q[\rho^\ast_{w_s}]=s$, and express $w_s$  as a function of $s$. While this is difficult for general values of $q$, $\xi$ and $m$, due to the complicated form of $\rho_w^\ast(\lambda)$, it may be done in  some cases.  In particular, fully analytic results can be obtained for $q=2$, $\xi=m=1/2$.  In this case, $I_2(s)$ can be written explicitly in phase II, displaying the simple form 
\be
I_2(s)=\frac{(s-\bar{s})^2}{2\gamma}\,,
\ee 
where $\bar{s}=3\ln 2-2\ln(1+\sqrt{2})$ is the average R\'enyi-$2$ entropy, while $\gamma\simeq 0.06$ is a numerical constant. Hence, for $w\in (w_1^\ast, w_2^\ast)$ the probability density for the R\'enyi-$2$ entropy is simply Gaussian. In phase I and III, instead, a large-$w$ expansion reveals that $I_2(s)$ develops logarithmic singularities for $s\to 0$ and $s\to\ln 2$: we find 
\be
I_2(s)=-\frac{1}{2}\ln |s-\tilde{s}|+ O(|s-\tilde{s}|)\,,
\ee 
with $\tilde{s}=0$,  $\tilde{s}=\ln 2$, respectively. 

We have tested our predictions against Monte Carlo simulations~\cite{krauth2006statistical}, numerically constructing a histogram of the probability $p_q(s)$ based on a sampling of~\eqref{eq:jacobi_spectrum}. Since the distribution of the R\'enyi entropies is highly peaked around its average, a standard Metropolis approach is not adequate to efficiently explore a wide range of its values, and we implemented the numerical scheme introduced in Ref.~\cite{nadal2011statistical}, where one forces the Metropolis algorithm to explore regions of large values of the R\'enyi entropy. As explained in~\cite{nadal2011statistical}, this method gives us  access to the derivative of the rate function $I^\prime_2(s)$ for finite systems~\cite{Note3}. The numerical data obtained using this method are reported in Fig.~\ref{fig:rate_renyi2} for the case $q=m^{-1}=2$, and different values  of $\xi$. The plot shows excellent agreement with our predictions, revealing that finite-size effects are  very small for the set of parameters considered.

\section{Conclusions}\label{sec:conclusions}
We have computed the large-deviation function for the entanglement of subsystems  in the steady state of the Q-SSEP. We have shown that its distribution is characterized by different phases connected by points where the probability density features singularities in its third derivative. Our work raises several questions. First, it would be interesting  to understand how our  predictions are modified for suitable generalizations of the model, such as the Q-SSEP with  dissipative boundaries~\cite{bernard2019open,bernard2020solution}, or its ``asymmetric'' version~\cite{jin2020stochastic}. Furthermore, a natural direction to explore pertains to the dynamics of entanglement, which should be in principle accessible from the stochastic equations of motion studied in~\cite{bauer2019equilibrium}. These questions are left for future work. 

\section*{Acknowledgments}
We are very grateful to Bruno Bertini and Satya N. Majumdar for reading the manuscript and for valuable comments. DB acknowledges useful discussions with M. Bauer and J.-B. Zuber. DB acknowledges support from CNRS, while LP acknowledges support from the Alexander von Humboldt foundation.

\bibliography{./bibliography}

\end{document}